\documentclass[twocolumn,showpacs,preprintnumbers,aps,amssymb]{revtex4-1}
\usepackage{amsmath}
\usepackage{natbib}
\usepackage[utf8x]{inputenc}
\usepackage[T1]{fontenc}
\usepackage[pdftex]{graphicx}
\usepackage{epstopdf}
\usepackage{color}
\usepackage{bm}
\usepackage{bigints}
\usepackage{relsize}
\usepackage{subeqnarray}

   \def\te{{}^3e} 

    \def\ham{\mathcal H}
\def\lag{\mathcal L}   
\def\pd{\partial} \def\da{\dot a} 
\def\f{\phi}   \def\a{\alpha}  
 \def\P{\varPi}    
  \def\t{\theta}   \def\L{\Lambda}
   
\newcommand{\dep}[2]{\frac{\pd #1}{\pd #2}}

 \def\dt{\dot t} \def\df{\dot \f}
\def\be{\begin{equation*}} \def\te{\end{equation*}} 
\def\bi{\begin{itemize}} \def\ti{\end{itemize}}
\def\ben{\begin{equation}} \def\ten{\end{equation}}

\begin{document}

\title{Anistropic Invariant FRW Cosmology}
\author{J. F. Chagoya}
\email{jfchagoya@fisica.ugto.mx}
\author{M. Sabido}
\email{msabido@fisica.ugto.mx}
\affiliation{Departamento  de F\'{\i}sica de la Universidad de Guanajuato,\\
 A.P. E-143, C.P. 37150, Le\'on, Guanajuato, M\'exico}%
\date{\today}
\begin{abstract}
In this paper we study the effects of including anisotropic scaling invariance in the minisuperspace
Lagrangian for a universe modelled by the Friedman-Robertson-Walker metric, a massless scalar field and
cosmological constant. We find that canonical quantization of this system leads to a Schroedinger
type equation, thus avoiding the \emph{frozen time} problem of the usual Wheeler-DeWitt equation.
Furthermore, we find numerical solutions for the classical equations of motion, and we also find evidence that under some conditions the big bang singularity is avoided in this model.  
\end{abstract}
\maketitle
\newpage
\section{Introduction}
The most widely accepted description of the universe is given by the $\Lambda$CDM model, which makes use of the Friedman-Robertson-Walker (FRW) metric in the framework of
General Relativity (GR) supplemented with a cosmological constant $\Lambda$. However, there are theoretical problems concerning the inclusion  of the cosmological constant (see \cite{Weinberg1989, Burgess2013, Padilla2015} for a review). {One of these problems is that} GR and Quantum Field Theory \textemdash the theory that describes the other fundamental interactions \textemdash are incompatible \cite{Goroff1986}. The tools of quantum field theory give a theory
with an ill ultraviolet (UV) behaviour when applied to GR. {This is one 
of the reasons for the 
interest in the search for alternative descriptions of gravity at high energies.}   Several modifications to GR have been proposed to describe gravitational 
physics at (pre-)inflationary and accelerated expansion epochs \cite[for a review of models of modified gravity see, for instance][]{citeulike:9404469}. 
 {The lack of a fundamental physical principle to construct
the ultraviolet theory of gravity makes the search for such a theory a complicated quest. A pragmatic approach is to consider GR} as the low energy limit of a more fundamental and so far unknown theory, the quantum theory of gravity. Any formulation of quantum gravity can have new symmetries and degrees of freedom in the UV regime, but  should recover or be compatible with GR in the low energy limit.

An UV completion to GR was proposed in \cite{Horava:2009uw} by making the theory invariant
under anisotropic scaling transformations. The resulting theory is power counting renormalizable, at the expense of losing Lorentz invariance in the UV limit as a consequence of the asymmetry  in the transformations that are enforced, 
\begin{equation*}t\to b^z t, \quad \vec{x}\to b \vec{x}.\end{equation*} 
The critical exponent $z$ is adjusted to have a renormalizable theory in the UV region, but in the IR flows to $z=1$ and Lorentz invariance is recovered. Although this theory has been considered as a real candidate for the UV region of GR, it is plagued by the existence of new degrees of freedom which make it incompatible with GR at low energies \cite{Sotiriou:2010wn}.

 The cosmological implications have been studied in Ho\v rava gravity by solving the equations of motion for the full theory (see \cite{Mukohyama:2010xz, Chakrabarti:2011hg}) and the quantum regime in \cite{octavio}.  In this paper we take a different approach,  we consider just 
one of the key ingredients of Ho\v rava's proposal, the anisotropic scaling invariance and apply it on a minisuperspace  having only two degrees of freedom corresponding to a metric function and a scalar field. This approach is similar to that of several works about minisuperspace noncommutativity \cite{Barbosa:2004kp} or quantum cosmology, a well established line of research that deals with quantization in the minisuperspace \cite{Misner:1969ae} and is likely to capture qualitative features of the full, {non-symmetry reduced} theory \cite{1993grg..conf...63H}. The resulting theory will be an effective theory,  that recovers  GR results when the anisotropy is eliminated.
For the minisuperspace model we choose the flat FRW metric, so we can easily contrast our results
with standard cosmology and the deviations obtained will be produced only by the modification to
GR due to the invariance under anisotropic scaling. We will construct an action for this  model  where the minisuperspace variables are compatible with the anisotropic transformations and time reparametrization invariance is preserved. We argue that the quantization of this model leads to a Schr\"odinger type equation for  $z\ne 1$, as we have shown for the Kantowski-Sachs model in \cite{Chagoya:2012tz}. This quantum dynamical equation with a first order time derivative allows us to construct a conserved probability current. In the case $z=1$, the system becomes singular and we return to the original WDW equation.
 We also obtain the classical equations of motion and find numerical solutions for different values of
the anisotropy index $z$. The analysis is done by comparing with the solution obtained in GR. 

The paper is organized as follows, in section \ref{sdos} the Lagrangian invariant under anisotropic scaling for the FRW model is presented as well as the symmetries of the theory. Then we use
canonical formalism to compute the Hamiltonian of the model and to obtain the modified dispersion relation and the Friedmann equations. In section \ref{stres} we present the exact solutions for the particular cases $\Lambda=0$ and $z=1$, and we analyze the numerical solutions for different values of the critical exponent $z$ -- also known as anisotropic index -- and the cosmological constant. Section \ref{scuatro} is devoted for conclusions and outlook.

\section{Anisotropic scaling invariant FRW cosmology}\label{sdos}

The flat Friedman-Robertson-Walker (FRW) cosmological model is described, in the ADM decomposition,
by the line element
\begin{equation}
 ds^2=-N^2dt^2+e^{2\a(t)}\left[dr^2+r^2(d\t^2+\sin^2\theta d\varphi^2)\right], \label{frw:metric}
\end{equation}
where $N(t)$ is the lapse function and $e^{\alpha(t)}=a(t)$ is the scale factor. The Einstein-Hilbert action minimally coupled to a massless 
scalar field $\phi$ and a cosmological
constant $\Lambda$ for this flat FRW metric is
\begin{equation}
S(\a,\f)=\bigintsss dt\left[-\frac{3a\da^2}{k^2N}+a^3\left(\frac{\df^2}{2N}+N\L\right)\right],
 \label{frw:accion}
\end{equation}
where the dot
represents derivatives with respect to $t$. 
{The first goal in this work is to look for generalizations of this action
incorporating the invariance under anisotropic
 scale transformations,
 $\vec x\rightarrow b\vec x$, $t\rightarrow b^zt$, where $z$  characterizes 
the anisotropy of the system. Besides this invariance, we also demand the action
 to be invariant under time reparametrizations, which means that if we
see the time coordinate as a function of a parameter $\tau$, then the action has to be
invariant under any choice of $\tau$. A final requirement that we impose on
the action is that in the limit $z=1$ it
has to reduce to the usual Lorentz invariant action for the FRW model in GR. The only action that
we could find satisfying the
 above conditions is}
\begin{align}
 S(a,\phi, t)&=\bigintsss{}{} d\tau\left\{-\frac{3a(\da^2)^z}{k^2N^z\dt^{z-1}}+
\frac{a^3(\df^2)^z}{2N^z\dt^{z-1}}\right. \nonumber \\
 &\qquad\qquad\quad+\left.\left(\frac{a^{4-z}N\L}{\dt}\right)^{\frac{z}{2z-1}}\dt\right\},
\label{frw:aniact}
\end{align}
where now dots denote derivatives with respect to 
 $\tau$. {The appearance of $\dt$ in the action is necessary to achieve explicit invariance
under time reparametrizations.} It is straightforward to see that when $z=1$ the standard FRW action is recovered.
  The invariances that we are requiring are satisfied if under time reparametrization $\tau=\tau(f)$ 
  we have $N\rightarrow\dot{\f}N$ and under anisotropic scaling
\begin{equation}\label{transfNL}
 N\rightarrow b^{\frac{3z-z^2+3}{z}}N, \hspace{1em} \L\rightarrow b^{\frac{z^2-6z-3}{z}}\L,
\hspace{1em} k^2\rightarrow b^{-2}k^2 .
\end{equation}

These transformations for $N$ might seem to restrictive, however 
we will see that this is not a problem when trying
to solve the equations of motion resulting from \eqref{frw:aniact}. 
Indeed, even the gauge $N=1$ that is usually chosen in order to simplify the equations 
in the $z=1$ case is compatible with \eqref{transfNL}.

Now we proceed to apply the canonical formalism to the anisotropic scaling 
invariant action \eqref{frw:aniact} in order
to obtain the respective Hamiltonian and the equations of motion.
\subsection{Canonical formalism for the anisotropic scaling invariant FRW action}
The dynamical variables that appear in action \eqref{frw:aniact}
 are the scale factor $a$, the scalar field $\f$ and {an explicit time parametrization  $t$}. Their corresponding canonical momenta are given, as usual, by
\begin{align}
 \P_a=\dep{\lag}{\da}&=-\frac{6az(\da^2)^{z-1}\da}{k^2N^z(\dt)^{z-1}}, \nonumber \\
 \P_\f=\dep{\lag}{\df}&=\frac{a^3z(\df^2)^{z-1}\df}{N^z(\dt)^{z-1}}, \nonumber \\
 \P_t=\dep{\lag}{\dt}&=-\frac{3a(\da^2)^z(1-z)}{k^2N^z(\dt)^z}+\frac{a^3(\df^2)^z(1-z)}
{2N^z(\dt)^z}\nonumber \\ 
&\hspace{1em}+\frac{1-z}{1-2z}a^{\frac{3z+z^2}{2z-1}-z}(N\L)^{\frac{z}{2z-1}}\dt^{\frac{z}{1-2z}},
\end{align}
where the Lagrangian $\lag$ is read off from the brackets in action \eqref{frw:aniact}.
We can try to construct a \emph{canonical Hamiltonian} $\ham_c$, but now not only we have the presence of 
the nondeterminate function $N$ but also we cannot write the attempted Hamiltonian
without explicit reference to the velocities of the canonical variables, this is a consequence
of the impossibility to invert the canonical momenta to have, e.g. $\dot{a}=\dot{a}(\P_a)$, so strictly speaking $\ham_c$
is an energy functional rather than a Hamiltonian. As already happens in GR we must interpret $\ham_c$ as a constraint
\begin{align}
\ham_c=&\da\P_a+\df\P_\f+\dt\P_t-\lag \nonumber \\
=&\frac{-3az(\da^2)^z}{k^2N^z(\dt)^{z-1}}+
\frac{a^3z(\df^2)^z}{2N^z(\dt)^{z-1}} \nonumber \\
&  +\frac{z}{1-2z}a^{\frac{3z+z^2}{2z-1}-z}(N\L)^{\frac{z}{2z-1}}\dt^{\frac{z}{1-2z}+1}\approx 0.\label{frw:hamaf}
\end{align}
{One can verify that by choosing $\dt=1$ then $\ham_c$ can be written as a true Hamiltonian
and it takes the form $N^{\frac{z}{2z-1}} \tilde\ham_c(a,\phi,\Pi_a,\P_f)$, making evident the roles
of $N$ as a Lagrange multiplier and of $\ham_c$ as a constraint.} 
The weak equality $\approx$ means that strictly $\ham_c$ is equal to zero only on a constraint hypersurface in the phase space. 
Comparing the last result with the expression for $\P_t$ we can obtain the formal relation
\begin{equation}\label{ht}
\ham_c=\frac{z}{1-z}\dt\P_t.
\end{equation}
Unlike the weak constraint \eqref{frw:hamaf}, this last relation is valid in the entire phase space. 
This is a result with remarkable consequences at the quantum level, since
the canonical quantization of this relation leads to a Schr\"odinger-type equation, where a first order derivative
in time appears \cite{Chagoya:2012tz}, {offering a possibility to solve the \emph{frozen time
 problem} of the Wheeler-DeWitt equation that appears when canonical quantization is 
applied to GR. For a review, see for example \cite{Isham2007}.} 
 However, our
focus in this work is on the classical theory.  We must note that the relation \eqref{ht}
between $\ham_c$ and $\P_t$ does not hold for the FRW model in GR since it is not
well defined for $z=1$. This should not be surprising
 as for $z=1$ the action \eqref{frw:aniact} does not contain any factors of
$\dt$, so a canonical momenta $\P_t$ cannot be defined.

Although we cannot write $\ham_c$ without any
explicit reference to the velocities of the dynamical variables, we can at least remove the dependence on $\da$ and $\df$.  It is straightforward to find
\begin{align}
\frac{z}{1-z}\dt\P_t=&-\left[\frac{3^{z-1}k^2N^z\P_a^{2z}}{a\dt^{1-z}12^zz}\right]^{\frac{1}{2z-1}}
+\frac{1}{2}\left[\frac{N^z \P_\f^{2z}}{a^3\dt^{1-z}z}\right]^{\frac{1}{2z-1}} \nonumber \\
&+\frac{z}{1-2z}\left[\frac{a^{4-z}N\L}{\dt}\right]^{\frac{z}{2z-1}}\dt. \label{frw:reldis}
\end{align}
{By construction, this model is invariant under time reparametrizations, thus we can choose $t=\tau$. This choice is convenient to cast the right hand side of \eqref{frw:reldis} as a true
Hamiltonian, which on-shell satisfies the weak constraint $\ham_c\approx 0$.
Let us now derive the Friedmann equations for this modified
FRW cosmological model.}
\subsection{Friedmann equations}
In the framework of general relativity it is customary to obtain the Friedmann equations 
through the Einstein equations. Nevertheless, they can also
be obtained from the Hamiltonian formalism as this is just another way to
describe the same physics. Specifically, they arise from the constraint $H=0$ and from the Hamilton 
equations for the momenta of the scale factor and of the scalar field. Using this identification,
we find that the analogues of the Friedmann equations for arbitrary values of $z$ are 
\begin{eqnarray}\label{fe}
3az(\dot a^2)^z-\frac{a^3}{2} z(\dot{\phi^2})^z-\frac{z}{1-2z}a^{\frac{4z-z^z}{2z-1}}
\Lambda^{\frac{z}{2z-1}}=&0,&\\
\vspace{1em}\nonumber \\
(6z+A(6z)^{\frac{2z}{2z-1}})(\dot a^2)^z+6z(2z-1)a (\dot a^2)^{z-1}\ddot a
 \nonumber \\
-Bz^{\frac{2z}{2z-1}}a^2(\dot\phi^2)^z -C\Lambda^{\frac{z}{2z-1}}a^{\frac{4z-z^z}{2z-1}-1}=&0,&\nonumber\\
\vspace{1em}\nonumber \\
3a^2\dot a  z (\dot \phi^2)^{\frac{2z-1}{2}}+a^3 z(2z-1)(\dot{\phi}^2)^{z-1}\ddot{\phi}=&0,&\nonumber
\end{eqnarray}
where we have defined the constants
\begin{align*}
A&=\frac{1}{1-2z}\left( \frac{3^{z-1}}{12^zz}\right)^{\frac{1}{2z-1}} , \hspace{1em}
B=\frac{3}{2(1-2z)}\left( \frac{1}{z}\right)^{\frac{1}{2z-1}}, \\
\vspace{1em}\nonumber \\
C&=\frac{z^z(4-z)}{(1-2z)(2z-1)}.
\end{align*}
We will refer to these generalized Friedmann equations simply as the Friedman equations.  
For $\Lambda\neq 0$, analytical solutions to \eqref{fe} cannot be found, so we need to resort to numerical methods.
Before showing numerical results we
present the exact solution for $\Lambda=0$ and we compare the respective numerical result 
with it, this is done in the next section. When $z=1$ the Friedmann
equations \eqref{fe} reduce to the ones obtained in GR for a
flat FRW metric, a cosmological constant and a massless scalar field:
\begin{align}\label{fegr}
3a \dot a^2+a^{3}\Lambda&=\frac{a^3}{2}\dot{\phi^2}\\
\dot a^{2}+2a \ddot a + \Lambda a^{2}&=
-\frac{1}{2}a^2\dot\phi^{2}\nonumber\\ 
3a^2\dot a \dot \phi+a^3\ddot{\phi}&=0.\nonumber
\end{align}
In the  next section we explore the properties of the model described by \eqref{fe}. 
\section{Numerical and analytical solutions}\label{stres}
We can find analytical solutions  for  $\Lambda=0$ and for $\Lambda\neq 0$ with $z=1$. 
Although some properties of the solutions for different $z$ can be studied in the first case, 
the main use of these exact solutions is that we can compare them to the numerical solutions
in order to check the consistency of our numerical results. 

For the case  $\Lambda=0$, we combine the first two equations in \eqref{fe} obtaining
\begin{equation}\label{fec}
2 {z} (\dot a  ^2 ) ^{z}+ (2z-1)^2 a (\dot a  ^2 ) ^{z-1}\ddot  a =0.
\end{equation}
The general solution to this differential equation is
\begin{equation}\label{solzgnol}
a(t)=\left[{\frac {(2 z-1) ^2 }{(4 {z}^2 +1-2z) (c_2+c_1 t) }}\right] ^{{\frac {(2 z-1) ^2 }{2 z-4 {z}^2 -1}}}.
\end{equation}
Once we have $a(t)$, the solution for $\phi$ is obtained directly from the third equation in \eqref{fe}, i.e. the momentum conservation
for the scalar field, but that is not relevant at this moment.
The quotient of $c_2$ and $c_1$ is related to the time $t_0$ at which $a=0$, whereas $c_2$ and $z$
determine the value of $a$ at $t=0$.
As expected, this solution is well behaved in the limit $z\to 1$ and it recovers the usual
 solutions $a(t)\sim t^{1/3}$ corresponding to the massless scalar field that we are considering. 
 Nevertheless, it is straightforward to see that \eqref{solzgnol} does not reflect 
the desirable characteristics that we would have expected in order
to make the addition of anisotropic invariance relevant at a
classical level, as would be the existence of an accelerating scale factor in the absence of the cosmological
constant or the removal of the singularity $a=0$. 

\begin{figure}
\includegraphics[width=0.4\textwidth]{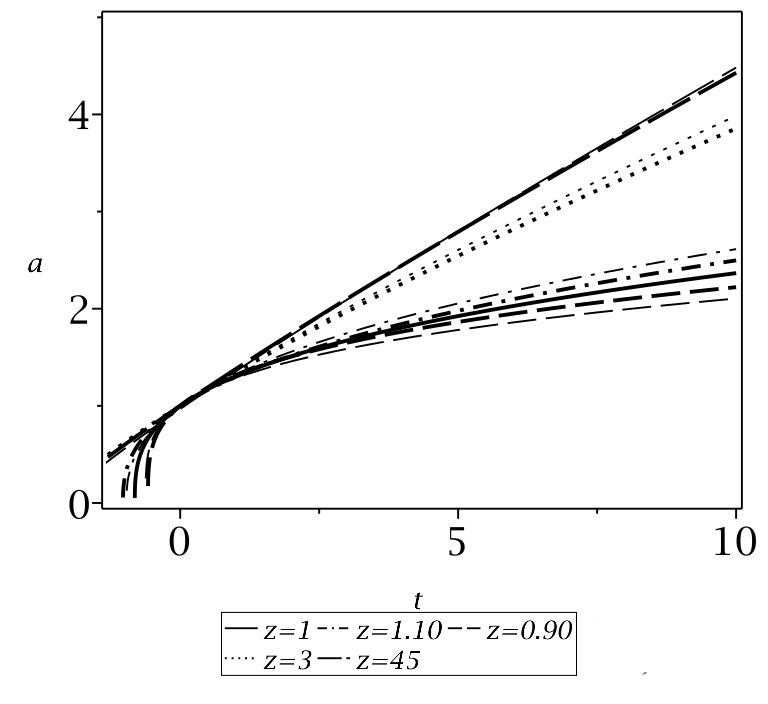}\caption{Solution to the Friedmann equations with
$\Lambda=0$, the same type of line is used both for the numerical and exact solutions that 
correspond to the same $z$, but the thicker lines are for the numerical results.}\label{L0}
\end{figure}

In Fig(\ref{L0}), we plot the analytical solutions  for some values of $z$ as well as the 
corresponding numerical solutions. The constants in \eqref{solzgnol} are chosen for each $z$ in
 order 
to make $\dot a|_{t=0}$ to coincide  for the numerical and exact solutions. The thick lines correspond to the analytical solutions and the thin lines to the numerical solutions for the same $z$. We can see that numerical numerical and exact solutions are in good agreement.  Also from this plot, we can verify  that for  $z>\frac{1}{2}$ the exponent in $a(t)$ is negative and therefore there is no acceleration. This can be verified explicitly from the analytical solution.
 
\begin{figure}
\includegraphics[width=0.4\textwidth]{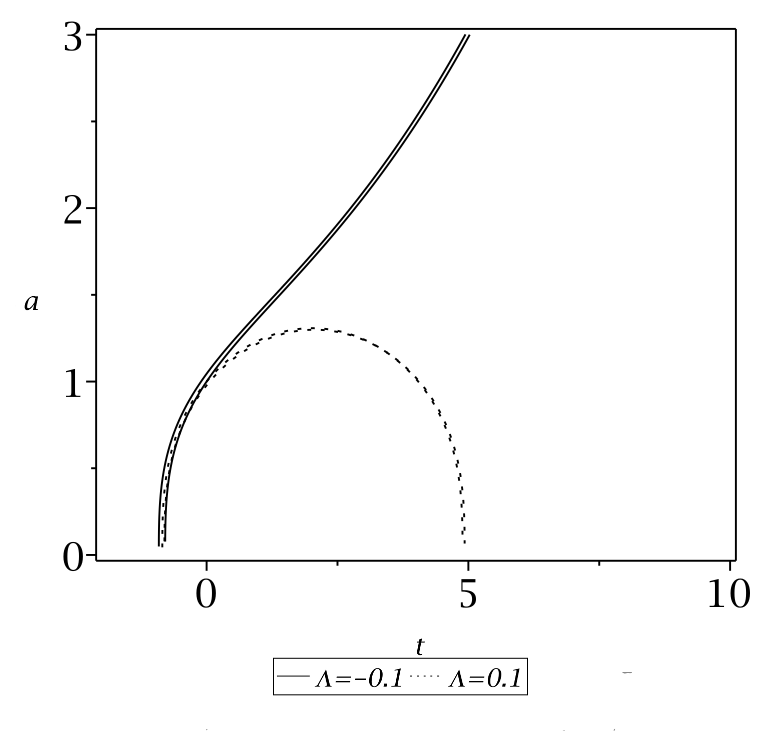}\caption{ Comparison of the numerical and analytical  solutions to the
Friedmann equations   for $\Lambda=-0.1$ and $\Lambda=0.1$. In both cases $z=1$.}\label{lpnz1}
\end{figure}

 For  the general case, $\Lambda\neq 0$ and $z \neq 1$ we will rely on the numerical results to extract conclusions on the model.
We take the system of Hamilton equations
derived from \eqref{frw:hamaf}, which has the advantage of being a first order system, and solve it numerically. The initial conditions
are chosen such that $H=0$. 
Before describing the results for $z\neq 1$, we check the consistency of the numerical results
for $z=1$ and $\Lambda$ positive or negative.  We do this by comparing the numerical results to the exact solution of the equation 
\begin{equation}
4 a \dot a  ^2 +2 a  ^2 \ddot  a +2 \Lambda a  ^3 =0,
\end{equation}
which is obtained by taking $z=1$ in the Friedman equations\eqref{fe} and making some
algebraic manipulations with them. A solution (not the most
general one) is given by
\begin{equation}
a(t)=\left[{-{\frac {C\sin(\sqrt 3 \sqrt {\Lambda}t+D) \sqrt 3 }{\sqrt {\Lambda}}}}\right]^{\frac{1}{3}}.\label{eq:az1}
\end{equation}
The constants $C$ and $D$ are related to the initial conditions for $a(t)$ and its velocity.
Figure (\ref{lpnz1}) shows a comparison of (\ref{eq:az1})  with the numerical solutions of the Hamilton's equations for $\L=0.1$ and $\L=-0.1$. In both cases there
exist a $t_0$  such that  $a(t_0)=0$, thus these solutions always have an initial big bang singularity. 
 
\begin{figure}
\includegraphics[width=0.4\textwidth]{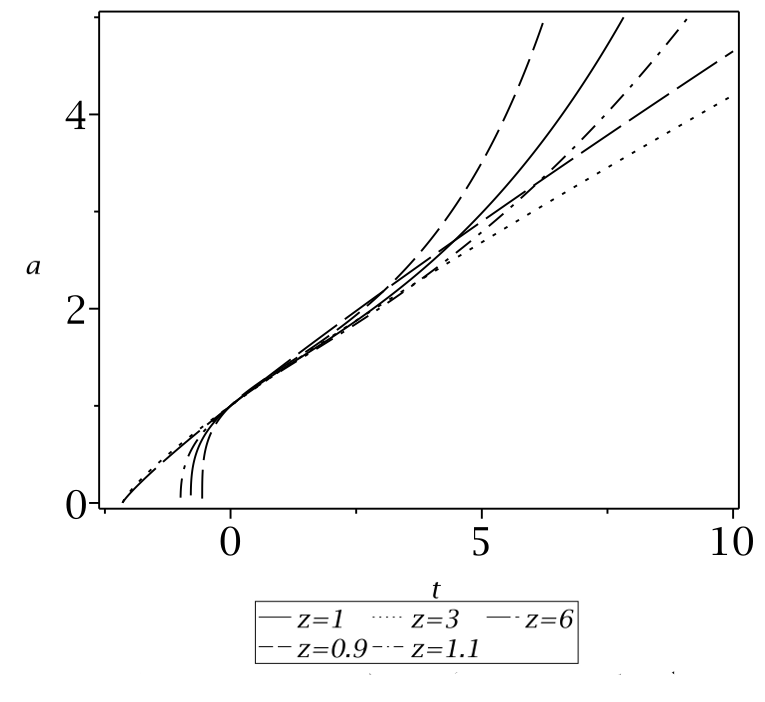}\caption{Numerical and exact solutions to the
Friedmann equations for $\Lambda=-0.1$ and different values of $z$.}\label{lambdaneg}
\end{figure}

Now we turn our attention to the solutions for $z\neq 1$. As already mentioned, here all the results are numerical.
Figure (\ref{lambdaneg})  shows the results for  $\Lambda<0$. We find that as $z$ becomes larger than 1 the accelerated growth of $a(t)$ is rapidly suppressed, in fact the scale factor
grows almost linearly with time for large $t$. In contrast,  when $z<1$ the accelerated growth of the scale factor is reinforced, so that the universe expands faster than in the solution for $z=1$.
There are no clear indications that the initial singularity can be avoided in this scenario. 

\begin{figure}
\includegraphics[width=0.4\textwidth]{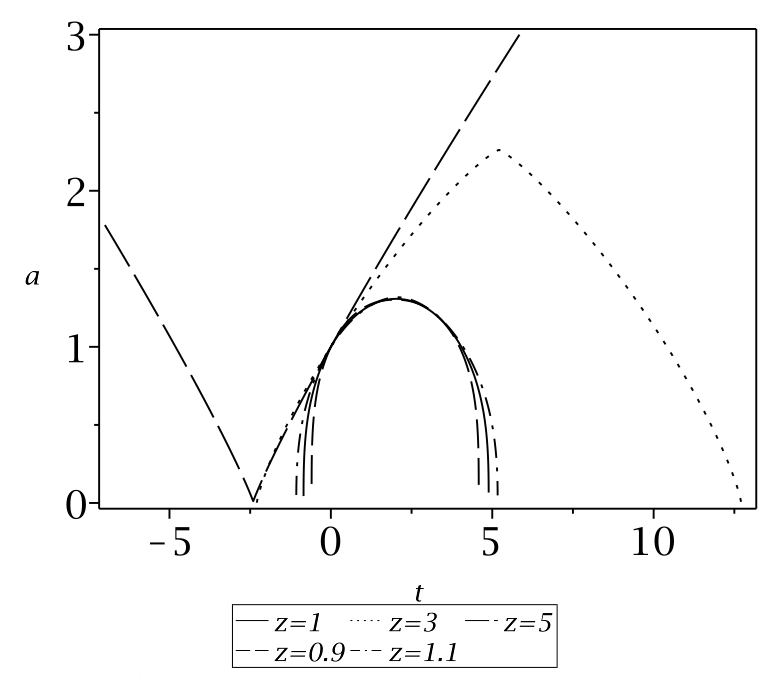}\caption{Numerical solutions to the
Friedmann equations for $\Lambda=0.1$ and different values of $z$.}\label{lambdapos}
\end{figure}

\begin{figure}
\includegraphics[width=0.4\textwidth]{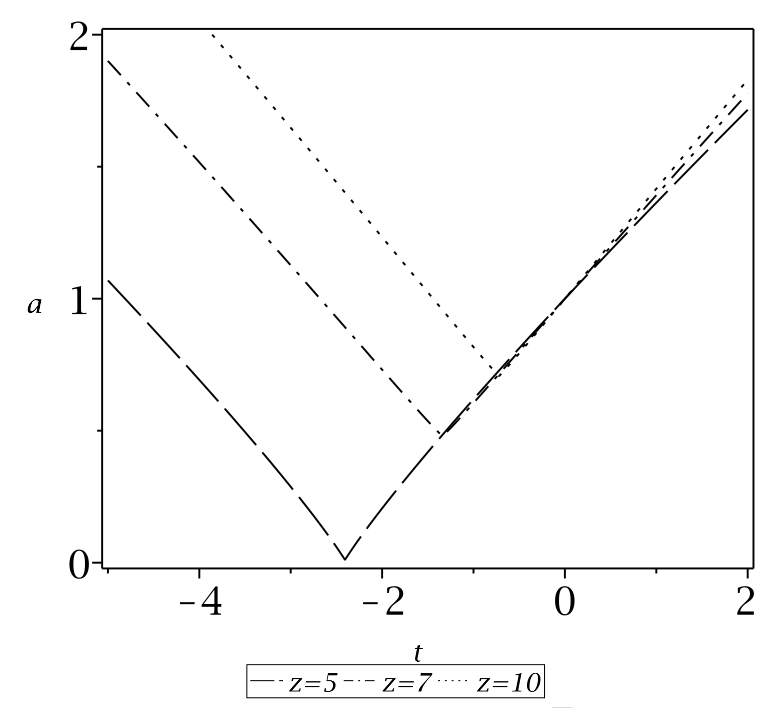}\caption{Numerical solutions to the
Friedmann equations for $\Lambda=0.1$. We see  a non zero minimal value for the scale factor.}\label{lambdaposzoom}
\end{figure}

For  $\Lambda>0$ the results are shown in Fig. \ref{lambdapos}. The value of $z$ has remarkable
consequences on
the behaviour of the scale factor. For example,  for $z=5$ we see that
there is not a $t_0$ such that $a(t_0)=0$. This characteristic is also present and more
noticeable for larger values of $z$, as shown in Fig.(\ref{lambdaposzoom}), where we can see in more detail the region where $a(t)$ reaches its minimum value, $a_{min}$. 
The precise value of $a_{min}$ depends on $z$ and  $\Lambda$. For $z< 5$, all the solutions
 describe
qualitatively the same type of universes, and we cannot say with confidence if all of them show a big bang singularity or not.
Unlike the case $\Lambda<0$, here there is not a relevant difference when $z$ changes from $z>1$ to $z<1$, epochs of accelerated expansion do not exist for any choice of $z$.
\section{Conclusions}\label{scuatro}
In this work we have analysed the effects of including invariance under
anisotropic scaling in the minisuperspace as an explicit symmetry of the action. We study a simple cosmological model
consisting of the flat FRW metric plus a massless scalar field and a cosmological
constant. A first relevant result of our analysis is that the Hamiltonian formulation of the model
reveals that the inclusion of anisotropic invariance is promising at the
quantum level, specifically with respect to the so called \emph{problem of frozen time},  since the  canonical formalism of the model leads
 to a dispersion relation that includes a linear momenta associated to the time
coordinate. After canonical quantization, this relation leads to a dynamical quantum equation that includes a first derivative in time, providing a way out to the frozen time problem.
 
In the main part of this work, we studied the classical solutions of the model
for various values of $z>\frac{1}{2}$. The reason to study only these values is the following:
for $z=\frac{1}{2}$ the last term in the action \eqref{frw:aniact} becomes singular, thus
this value of $z$ can be
thought as separating the anisotropic index in the ranges $z>\frac{1}{2}$, which 
includes the usual Lorentz invariant solutions when $z=1$, and $z<\frac{1}{2}$.
{Although we can suppose that the anisotropic invariance was present at very early 
stages of the universe and therefore $z$ could take any value, we want to be able to recover 
the Lorentz invariant ($z=1$) solutions at
late times. Assuming that the transition from anisotropic to Lorentz invariance is continuous, this requirement imposes that we have to stay in the $z>\frac{1}{2}$ branch.}
 
With respect to the solutions that we found to the anisotropic scaling invariant model,
 some remarks are in order:
 \begin{itemize}
 \item The limit $z\to 1$ is well defined for all of them.
 \item Without a cosmological constant, the inclusion of anisotropic invariance
 makes the volume of the universe to grow almost linearly with time. When this 
 linear regime is reached the velocity $\dot a$ is larger for larger $z$.
 \item With a negative cosmological constant, the accelerated growth of $a(t)$ 
 is suppressed for $z>1$, whereas for $z<1$ it is reinforced. 
 \item With a positive cosmological constant we have the most remarkable modification
 introduced by the invariance under anisotropic scaling, namely, we can remove the Big Bang singularity, obtaining
an $a_{min}\neq 0$. The minimum $z$ for which this is attainable depends on $\Lambda$ and on the initial conditions. 
 \end{itemize}
 As an overall conclusion, the inclusion of anisotropic invariance results in desirable characteristics both at the quantum and classical levels. It gives a possible 
 solution to the problem of time in quantum cosmology as well a resolution of the Big Bang singularity.
{However, we need to seek further for explicit exact solutions that allow us to establish firmly
the existence of a non-vanishing minimum value for the scale factor, and to find analytical conditions to determine the relation between the values of $z$, $a_{min}$ and
$\Lambda$. The existence of a mechanism to change dynamically the anisotropic index in such
a way that the model can flow from an anisotropic invariant epoch at high energies -- close to the Planck scale --
to a Lorentz invariant epoch at lower energies is a non-trivial open question that would
be interesting to study, since such a mechanism would allow us to have relevant deviations from classical solutions at high energies while recovering standard general relativity results in the infrared regime.}
\section*{Acknowledgments}
This work is  supported by   CONACYT grants 167335, 179208 and DAIP640/2015 and is part of the PROMEP research network  ``Gravitaci\'on y F\'isica Matem\'atica".
\end{document}